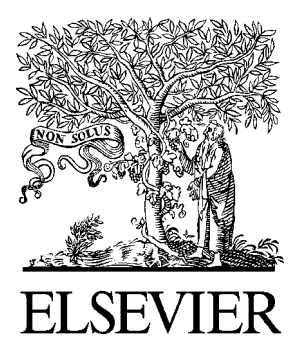

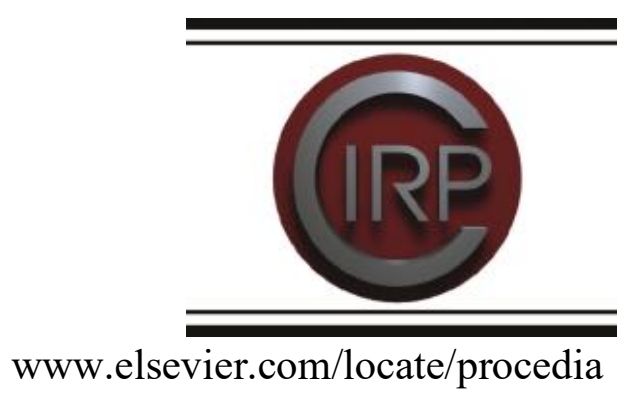

# Hemispherical Ray-Casting Analysis for Milling Configuration Classification and Machinability Assessment

Serafeim Baltadouros[a,*], Bert Lauwers[a], Joost Duflou[a]

[a] KU Leuven, Department of Mechanical Engineering, Division of Manufacturing Processes and Systems (MaPS) and Flanders Make@KU Leuven · M&A, 3001 Leuven, Belgium

* Corresponding author. Tel.: +30-694-585-9222. *E-mail address:* serafeim.baltadouros@kuleuven.be

**Abstract**

Early-stage manufacturability assessment reduces design iterations and costly downstream changes. This paper presents a GPU-accelerated geometric analysis method that evaluates CNC machinability and setup requirements directly from CAD files. A multi-stage pipeline uses hemispherical ray-casting to quantify surface accessibility, followed by directional variability analysis to extract dominant machining axes. Threshold-based classification determines 3-axis versus multi-axis machining. For 3-axis parts, orthogonal ray-casting and reachability optimization compute the minimum setup count and explicitly flag inaccessible surfaces. Validation cases show computation times sufficiently short to support interactive CAD workflows, enabling quantitative manufacturability feedback during the design stage without requiring manufacturing expertise.

## 1. Introduction

The integration of manufacturing considerations into the early stages of product design represents a critical challenge in contemporary engineering practice. Designers require immediate feedback regarding the manufacturing implications of their design decisions, particularly concerning the feasibility of producing components using specific machining processes. This need is especially pronounced in the context of CNC milling operations, where geometric constraints directly determine machining configuration requirements that affect manufacturability effort, production time and cost.

Current industrial practice relies heavily on Computer-Aided Manufacturing (CAM) software to evaluate the machinability of designed components [1]. However, this approach presents significant limitations for design-stage decision-making. The conventional workflow requires a finalized design to be parsed through CAM systems, where process planning algorithms generate toolpaths and identify potential manufacturing issues [2]. This sequential process is inherently time-consuming and interactive, often requiring skilled manufacturing engineers to interpret CAD models and configure machining simulations for design validation [3]. The feedback loop between design and manufacturing thus occurs late in the development cycle, when design changes become increasingly expensive and disruptive.

The fundamental problem lies in the disconnection between Computer-Aided Design (CAD) environments, where designers work iteratively on geometric models, and the manufacturability analysis that occurs downstream in dedicated CAM systems. Designers make critical decisions about form, features, and configurations without real-time awareness of how these choices impact manufacturing feasibility and cost [4]. This separation creates a critical gap: by the time manufacturing issues are identified, the design may be too far advanced for efficient modification, leading to costly engineering change requests and project delays [1].

Recent research has begun to address this limitation by developing methods for rapid manufacturability assessment that can operate within CAD environments or alongside them [1,2]. These approaches aim to provide immediate visual and numeric feedback about manufacturing constraints, enabling designers to make informed decisions without waiting for formal manufacturability reviews. This research extends the integration of manufacturing feedback into the design environment by developing real-time accessibility analysis specifically for milling operations. The proposed approach provides designers with immediate insight into two critical manufacturing implications of their design decisions: (1) machining configuration requirements, determining whether a part is feasible for 3-axis milling or necessitates multi-axis CNC, and (2) setup complexity, identifying the minimum number of setups required to access all machinable surfaces, thereby providing an early indicator of manufacturing effort and cost. Specifically, this paper introduces a geometric analysis pipeline that evaluates surface accessibility, identifies dominant machining directions, classifies machining configuration requirements, and determines the minimum number of setups for 3-axis parts. By making these capabilities available at the design stage, the proposed approach seeks to bridge the gap between geometric design decisions and their downstream manufacturing consequences. To contextualize this contribution, the following sections review existing approaches to tool accessibility analysis and setup planning, examining both their capabilities and limitations in supporting early-stage design decision-making.

## 2. State of the Art

Geometric accessibility analysis has been an active area of research for many years, with most contributions built on surface-based models. Early contributions established the hemisphere as the fundamental representation of a surface point's visibility range, such that the full visibility map of a surface is derived from the intersection of all individual point hemispheres [5,6]. Accessibility was subsequently categorized into local and global forms [7]. Local analysis considers only nearby obstacles, while global analysis also includes external elements such as fixtures or clamps. To handle complex surface representations, the convex hull property of NURBS control polygons was exploited to detect global tool interference [8]. The accessibility map and configuration-space (C-space) formulations were also introduced. The accessibility map records feasible tool postures across a surface to guide tool path computation and avoid collision while ensuring smooth posture transitions [9]. C-space approaches explicitly map all tool positions and orientations into a unified space where collisions can be identified, enabling gouge-free path generation [10]. GPU rendering capabilities have also been exploited across several accessibility analysis approaches, such as the computation of accessibility cones for determining valid cutter postures[11]. In parallel, ray-casting has long been a foundational technique in computer graphics [12], applied in CAD rendering pipelines and photo-realistic simulation [13,14]. Its use in machining has remained limited in computing discrete ray-surface intersections [15], constrained by the computational cost of the time and limited to simple geometries. Building on these advances, the present work employs ray-casting techniques to generate manufacturability indicators during the early design phase.

Research on setup planning has developed alongside accessibility analysis. Broadly, these methods fall into two main categories: feature-based approaches, which rely on the identification of machining features to guide tool selection and orientation, and feature-free approaches, which operate directly on the geometry without explicit feature recognition, enabling greater flexibility for complex and freeform surfaces.

Feature-based setup planning identifies geometric features and assigns machining operations and tool approach directions [16]. These methods integrate manufacturing knowledge and support automated process planning [17,18], fixture planning [19,20], and optimization strategies. However, feature-based methods fail for freeform and generative geometries, require manual rule updates, and are unsuitable for dynamic manufacturing environments. Feature-free methods operate directly on geometric models without feature recognition [21–23]. They analyze surface normals, curvature, and accessibility using slicing and visibility mapping. These approaches handle arbitrary freeform geometries but sacrifice some manufacturing knowledge integration. While they demonstrate feasibility of geometry-based setup planning, they have not been integrated into rapid design-stage assessment tools yet. The integration of accessibility metrics with setup determination to produce actionable design-stage metrics in real time, remains an open challenge.

Building upon these limitations and identified research gaps, this paper advances a feature-free, geometry-driven approach aimed at enabling real-time manufacturability assessment during the early design stage. First, it automatically determines required machining axis configurations based on surface accessibility. Second, it proposes directional ray-casting analysis, recognizing occluded areas on the design, and establishing the foundation for setup evaluation. Third, it establishes setup count as a quantitative manufacturability complexity metric, providing an intuitive proxy for manufacturing cost and difficulty.

The feature-free framework operates directly on STEP/STL geometry, enabling analysis on feature-free designs. GPU acceleration ensures computational efficiency, demonstrating feasibility for future real-time CAD integration.

## 3. Methodology

Manufacturing complexity for milled components is influenced by numerous interrelated factors that must be resolved during process planning. Among these, two critical considerations significantly impact production feasibility and cost: the required machine configuration and the number of setups needed to access all machinable surfaces. The proposed methodology enables early-stage geometric assessment of machine requirements and setup complexity, providing designers with immediate manufacturability feedback and reducing planning effort and redesign risk.

The distinction between 3-axis and multi-axis machining represents a fundamental capability and cost threshold. 3-axis machines restrict tool access to fixed directions, making them suitable for prismatic parts but requiring multiple setups for complex orientations. Multi-axis machines provide additional rotational degrees of freedom, enabling access to complex surfaces with fewer setups but at increased equipment, programming, and operational complexity.

For 3-axis-suitable parts, the number of required setups directly affects manufacturing effort, fixture design, alignment accuracy, and error accumulation. Minimum setup count therefore serves as an intuitive quantitative metric for manufacturing difficulty and enables comparative evaluation of design alternatives prior to detailed process planning.

The proposed pipeline evaluates these indicators through multi-stage geometric analysis on triangulated meshes. Hemispherical ray-casting quantifies surface accessibility, followed by variability analysis to classify parts as 3-axis or multi-axis. For 3-axis parts, directional accessibility analysis identifies feasible tool approach directions, which are combined with tool reachability constraints to compute the minimum setup count via constrained optimization.

All computations are GPU-accelerated using PyTorch and NVIDIA Warp, enabling efficient analysis of high-resolution meshes. The following subsections describe the methodology implementation, and derived metrics.

### 3.1. Hemispherical Accessibility Analysis

Hemispherical accessibility analysis quantifies the geometric openness of each surface element by evaluating the proportion of directions from which the surface is visible without occlusion. The analysis operates on a triangulated mesh extracted from the CAD file. For each triangular face, the centroid, surface normal, and area are computed.

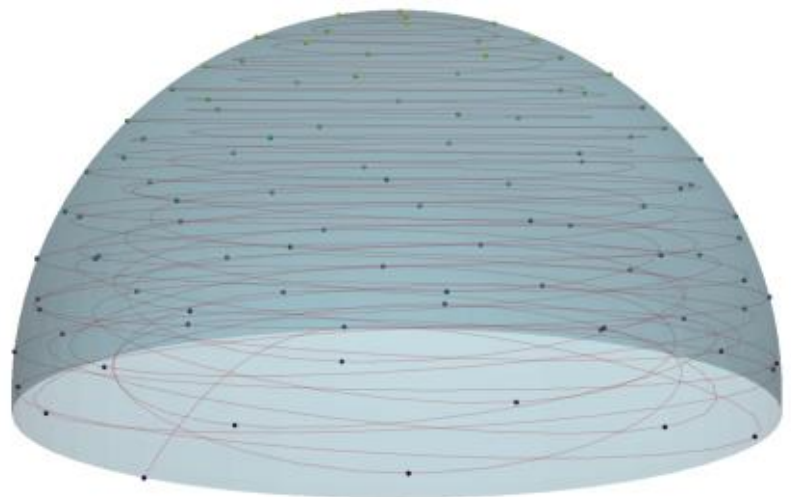

Fig. 1. Sample points generated on a hemisphere.

Accessibility is evaluated by casting rays over a uniformly sampled hemisphere aligned with the local surface normal (Fig. 1). The hemisphere samples are rotated to align with each face normal, and rays are cast from a slightly offset centroid to avoid self-intersection. GPU-accelerated ray-triangle intersection tests determine whether rays are obstructed by geometry (Fig. 2). Rays colliding with the surface are excluded.

For each face, the analysis yields the number of unobstructed rays and an accessibility ratio. An area-weighted overall accessibility score is computed across the entire part. The spatial distribution of unobstructed rays provides directional information that is later used to assess machining axis requirements.

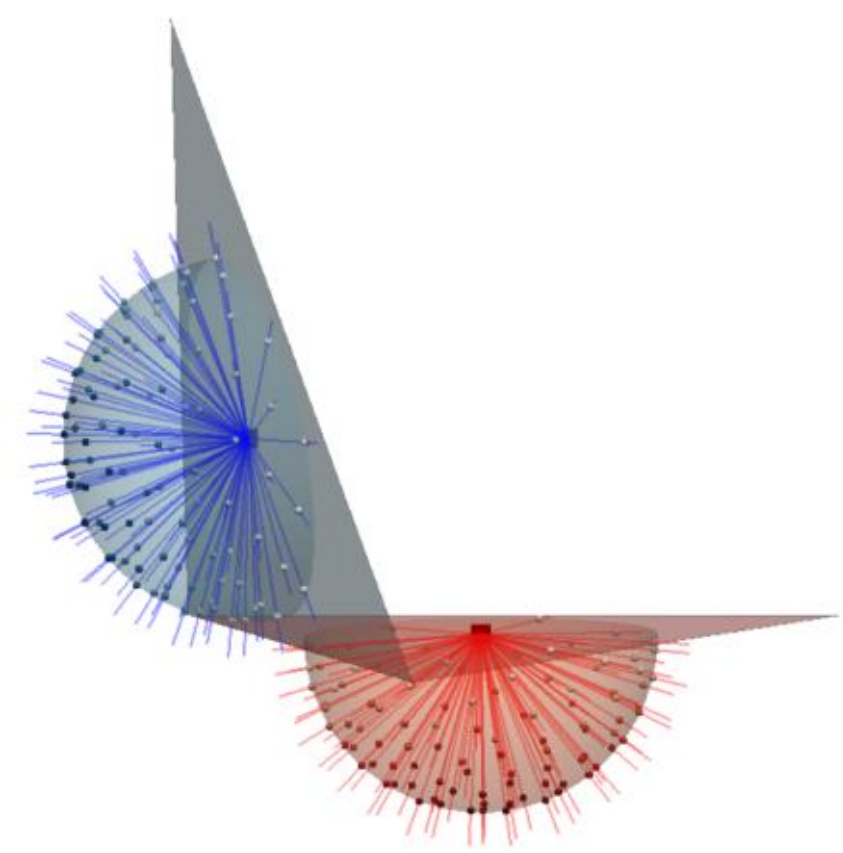

Fig. 2. Ray casting from facet centroid towards sample points directions.

### 3.2. Directional Accessibility Analysis

Directional accessibility analysis identifies which principal machining directions can access each surface without collision. This stage is applied only to parts classified as potentially suitable for 3-axis machining. Prior to analysis, each part is reoriented into a canonical machining frame consistent with how it would be fixtured from stock material, establishing six principal directions in a physically meaningful reference frame. Rays are then cast along each of these directions from an offset centroid, tested for intersection, using the same GPU-accelerated kernel.

The result is a binary direction accessibility matrix indicating whether each face is accessible from each principal direction, along with the number of accessible directions per face. Faces with no accessible direction indicate features that cannot be machined using standard 3-axis setups. These data are later used to compute the minimum setup count.

### 3.3. Manufacturability Indicators Evaluation

This stage integrates metrics derived from both hemispherical and directional analyses to determine whether 3-axis or multi-axis machining is necessary and to establish the minimum number of setups required for 3-axis parts.

Directional biases in unobstructed rays are analyzed to identify dominant axis alignment and directional concentration for each face. Three indicators are used to classify machining requirements: distribution of dominant axes across faces, mean directional concentration, and overall accessibility. Thresholds for these indicators are empirically determined using benchmark parts. If any threshold is exceeded, the part is classified as requiring multi-axis machining; otherwise, the pipeline proceeds to setup planning evaluation. This classification is purely geometry-driven, as process-level constraints are outside the current scope.

Classification indicators are defined as follows. Overall accessibility is the average ratio of unobstructed to total rays across all part faces, reflecting how much of the part surface is geometrically reachable from principal directions. Dominant axis distribution reports what percentage of faces have their largest mean unobstructed ray component along each principal axis. The three percentages always sum to 100% by construction, where a uniform distribution across all three axes reflects spherical-like geometry typically requiring multi-axis machining, while a strong concentration toward one or two axes suggests geometry more suited to 3-axis configurations. Mean axis concentration measures how exclusively each face is accessible from its dominant axis, as opposed to being reachable from multiple directions. For example, a deep hole concentrates rays along its axis, while a flat face receives rays from multiple directions, resulting in lower concentration. The mean is averaged over all faces.

For 3-axis parts, accessibility is combined with tool reachability constraints. For each accessible direction, the tool travel distance is evaluated against a user-defined threshold representing the effective cutting tool length available in the user's tooling setup. Directions exceeding this threshold are discarded. Bounding box faces are excluded as fixture surfaces. Remaining faces are classified as viable, over-threshold, or fully inaccessible. Setup planning is performed on viable faces using a greedy set-cover strategy that selects the minimum number of setup orientations required to cover all machinable faces. The resulting setup count serves as a quantitative manufacturability complexity metric, while over-threshold and inaccessible faces are flagged for design review. The full pipeline provides early-stage feedback on machining requirements and manufacturing complexity, which is validated in the subsequent case studies.

The following section validates this pipeline through representative case studies demonstrating its application to parts with varying geometric characteristics.

## 4. Case Study Validation

The proposed methodology is validated using two representative case studies[24,25]. The selected parts represent diverse geometric scenarios: a complex multi-axis component and a prismatic part suitable for 3-axis machining. These cases test both machining classification and setup planning capabilities. Analysis results are presented here, with evaluation discussed in the following section.

### 4.1. First Case

The first case study analyzes an impeller, a canonical multi-axis machining example due to its continuously curved surface. Hemispherical accessibility results (Fig. 3) show strong occlusion in blade passages and the hub region, with higher accessibility on outer surfaces.

Quantitative results are summarized in Table 1. The overall accessibility of 49.9% indicates significant self-occlusion. Dominant axis distribution is nearly uniform across X, Y, and Z directions, and the mean axis concentration is 0.419. These metrics exceed the classification thresholds, leading to correct identification of multi-axis machining requirements. This classification aligns with industrial practice for impeller manufacturing.

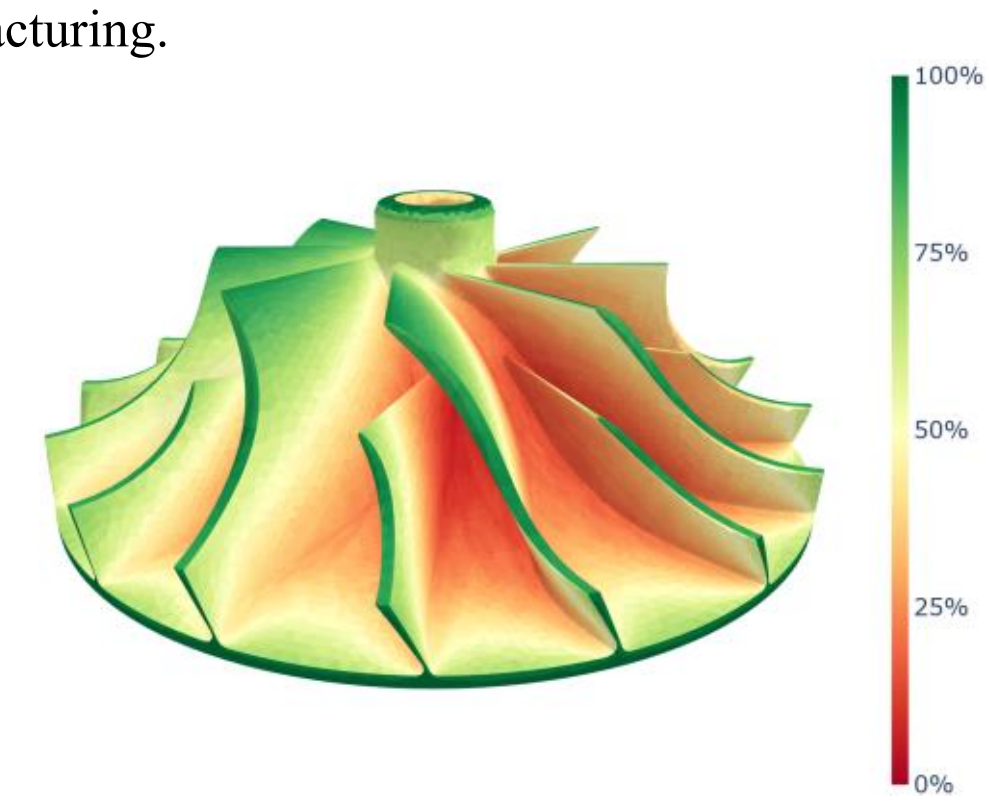

Fig. 3. Hemispherical accessibility analysis results - Case study 1.

Table 1. Manufacturing Complexity Assessment Summary – Case study 1.

| Metric | Value |
|---|---|
| Overall Accessibility | 49.9% |
| Dominant X-axis Faces | 34.4% |
| Dominant Y-axis Faces | 30.9% |
| Dominant Z-axis Faces | 34.7% |
| Mean Axis Concentration | 0.419 |
| Machining Requirement | Multi-axis |

### 4.2. Second Case

The second case study analyzes a prismatic housing component with limited directional bias. Hemispherical accessibility results (Fig. 4) show high overall accessibility of 72.9% (Table 2). Directional variability reveals strong Z-axis dominance, clearly satisfying 3-axis classification criteria.

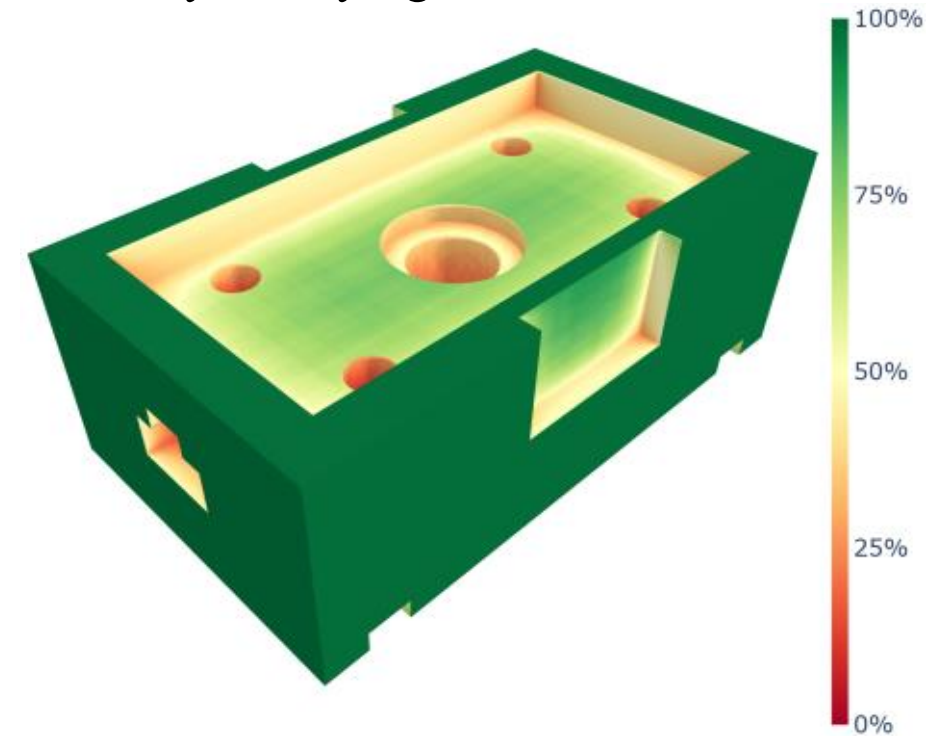

Fig. 4. Hemispherical accessibility analysis results - Case study 2.

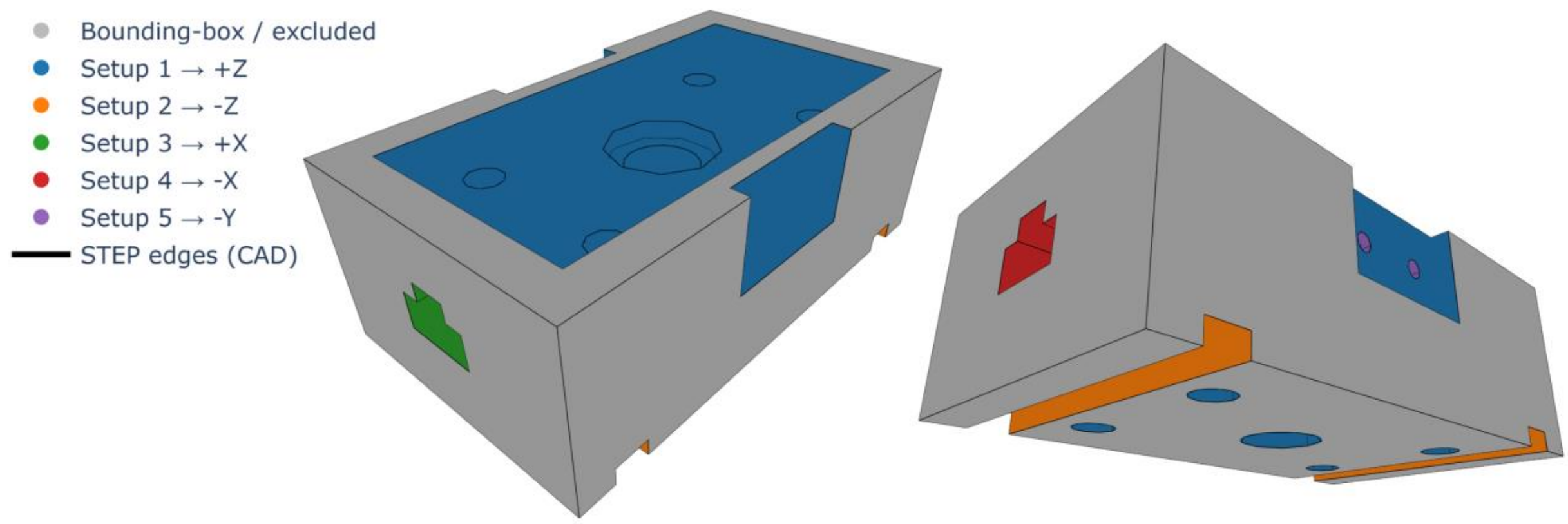

Fig. 5. Minimum setup count evaluation - Case study 2.

Directional accessibility and setup planning (Fig. 5) indicate that five setups are required to access all viable faces, with most operations aligned to the dominant axis and supplementary setups for side features.

Table 2. Manufacturing Complexity Assessment Summary – Case study 2.

| Metric | Value |
|---|---|
| Overall Accessibility | 72.9% |
| Dominant X-axis Faces | 16.3% |
| Dominant Y-axis Faces | 25.7% |
| Dominant Z-axis Faces | 58.0% |
| Mean Axis Concentration | 0.393 |
| Machining Requirement | 3-axis |

Table 3 summarizes computational performance metrics. GPU-accelerated analysis enables processing of high-resolution meshes within sufficiently short runtimes, demonstrating computational efficiency and suitability for early-stage design assessment.

Table 3. Computational performance metrics for case study validation.

| Metric | Case 1 | Case 2 |
|---|---|---|
| STEP File Size (kB) | 2703 | 67 |
| Mesh Face Count | 123958 | 22530 |
| Hemisphere Sample Points | 150 | 150 |
| Computation Time (s) | 4.50 | 0.18 |

## 5. Discussion

The case study results demonstrate the methodology's capability to differentiate machining requirements and quantify manufacturing complexity across diverse geometric scenarios through purely geometric analysis.

The impeller (Case 1) presents the characteristic complexity of multi-axis components: balanced dominant axis distribution and low overall accessibility. The near-equal partitioning of faces across all three axes indicates no clear primary machining direction, while the low accessibility score reflects significant self-occlusion from the curved blade passages. The classification of multi-axis machining requirements aligns with industrial practice where impellers require 5-axis simultaneous machining to access the compound curvature of their surfaces without workpiece repositioning. The hemispherical analysis successfully captures the geometric signature of multi-axis requirements without requiring feature recognition or manufacturing knowledge.

The housing component (Case 2) demonstrates strong Z-axis dominance and high overall accessibility, reflecting its simplistic geometric configuration with primarily top-down oriented features. The 5-setup minimum represents efficient manufacturing planning: primary operations from the Z-direction access the dominant surfaces, while supplementary X and Y setups address side features. A process planner might potentially select 6 setups if surface finish requirements on specific side features necessitate optimized tool engagement angles or reduced tool deflection. However, the framework correctly identifies 5 as the theoretical minimum required for geometric accessibility under the imposed distance constraints, providing designers with the lower bound for manufacturing complexity.

Computational results demonstrate the feasibility of integrating the analysis into design workflows. All case studies, involving tens of thousands of faces and millions of ray queries, completed within seconds using GPU acceleration (with the system specifications detailed in Table 4). This performance supports iterative design refinement with immediate manufacturability feedback, in contrast to traditional workflows where manufacturing evaluation occurs only after design finalization.

Table 4. System specifications for computational benchmarking.

| Hardware Component | Specification |
|---|---|
| CPU | AMD Ryzen 7 5800H |
| GPU | NVIDIA GeForce GTX 1650 |
| RAM | 15.3 GB |
| CUDA Version | 13.0 |
| Python Version | 3.9.23 |

## 6. Conclusion and Future Work

This research presents a GPU-accelerated geometric analysis framework for automated machinability assessment and manufacturing complexity directly from CAD files. The methodology addresses a key gap in design-stage manufacturability evaluation by providing early feedback on machine configuration requirements and quantitative complexity indicators prior to process planning.

The multi-stage pipeline converts geometric metrics into actionable manufacturing insights. Validation on two representative case studies demonstrates correct classification of machining requirements and accurate minimum setup count estimation consistent with process planning decisions. The feature-free approach operates on arbitrary triangulated geometry and handles complex freeform surfaces without relying on feature recognition or manufacturing knowledge databases, supporting early-stage design decision-making.

Future work could extend the framework's capabilities in several directions. A critical enhancement involves automatic identification of datum surfaces for each setup orientation, enabling proper geometric dimensioning and tolerancing analysis by establishing reference frames for each machining operation. Additionally, reconstruction of intermediate part geometries between setups would provide crucial information for 3-axis toolpath generation, where stock material remaining from previous operations affects subsequent machining strategies and tool accessibility. These extensions would further bridge the gap between design-stage analysis and detailed process planning, enabling more comprehensive manufacturability assessment within interactive design workflows.